\documentclass[final]
  {aipproc}

\usepackage{amsmath,bm}

\layoutstyle{6x9}

\newcommand{\bdf}{\mathbf{f}}
\newcommand{\ba}{\mathbf{a}}
\newcommand{\bdm}{\mathbf{m}}

\def\dblone{\hbox{$\bm{1}\hskip -1.2pt\vrule depth 0pt height 1.6ex width 0.7pt
                  \vrule depth 0pt height 0.3pt width 0.12em$}}

\begin{document}

\title{Renormalization Group Analysis of Supersymmetric Particle Interactions}


\author{Andrew D. Box}{
  address={Department of Physics and Astronomy, University of Oklahoma, Norman, OK 73019, USA}
}

\begin{abstract}
We reexamine the renormalization group equations (RGEs) for the dimensionless and dimensionful parameters of the Minimal Supersymmetric Standard Model (MSSM), incorporating 1-loop thresholds. The inclusion of these thresholds necessarily results in splitting between dimensionless couplings which are equal at the tree level. Assuming that the SUSY-breaking mechanism does not introduce new intergenerational couplings, we present the most general form for high-scale, soft-SUSY-breaking (SSB) parameters. With this as our boundary condition, we consider illustrative examples of numerical solutions to the RGEs. In a supersymmetric grand unified theory with the scale of SUSY scalars split from that of gauginos and higgsinos, we find that the gaugino mass unification relation may be violated to the order of 10\%. Further, we consider the rate for the flavor violating decay of the lightest stop to charm plus neutralino. We find that using the complete RGE solution as opposed to the commonly used `single-step' integration of the RGEs can qualitatively change the picture of event-topologies from top-squark pair production, or from gluino production if gluino to stop plus top is the dominant gluino decay mode.
\end{abstract}

\maketitle


\section{Introduction}

The MSSM contains a dazzling number of free parameters, which are largely a result of our ignorance of the mechanism of supersymmetry breaking. To counter this issue, most supersymmetric models reduce the parameters to a manageable number by way of high scale ans\"{a}tze. Starting from these high scale inputs, the weak scale parameters of the theory can be calculated via the renormalization group equations (RGEs).

For our predictions to be accurate to two-loops, we must take full account of the non-degenerate SUSY mass spectrum by introducing particle thresholds into the one-loop RGEs. Once the various particles are decoupled from the theory, we must also take account of splitting between couplings that are equal in the SUSY limit \cite{RGE1}.

The RGEs are constucted to describe a collection of effective theories with varying particle content that are valid at different scales. Following Ref.~\cite{CPR}, we make the assumption that each particle with mass $M_i$ is included in the effective theory only if $Q>M_i$. In this manner the $\beta$-functions are as in the MSSM at a scale above the masses of all SUSY particles. Moving down in scale, the particles are decoupled individually until we have only SM particles in the theory and the $\beta$-functions are those of the SM.

If we have an appropriate theory of flavor at the high scale, we can use this method to obtain predictions for the level of flavor violation at the weak scale. To illustate the importance these effects we obtain the rate for the two-body flavor changing decay of the top squark, using the RGEs to calculate the mixing between top and charm squarks.

\section{Particle Decoupling and Solutions to the RGEs}

Our starting points for including threshold effects in the MSSM RGEs are the general results in Ref.~\cite{MVLuo} for the dimensionless and dimensionful couplings. We have converted their work to apply to four-component fermions and complex scalars, and obtained the full system of threshold RGEs for the gauge couplings and Yukawas \cite{RGE1}, as well as $\mu$, the gaugino masses, the soft masses and trilinear SSB parameters \cite{RGE2}.

We include particle threshold effects in the RGEs for SSB parameters, being careful to freeze the running as the scale, $Q$, crosses each eigenvalue, thereby removing the contribution that this mass eigenstate makes to the overall running. In finding eigenvalues (solely for the purpose of identifying the location of the thresholds) we neglect the left--right mixing between sfermions, a sensible approximation when sfermion SSB parameters are larger than the weak scale.

Our high scale boundary conditions are the most general form for SSB matrices that does not introduce new sources of flavor violation. In addition, we include arbitrary matrices, $\mathbf{T}_{\{Q,L,U,D,E\}}$ and $\mathbf{Z}_{\{u,d,e\}}$, that allow for the introduction of dependence of the theory on all quark rotation matrices, as opposed to just the KM.

We thus parametrize the SSB sfermion mass and ${\bf a}$-parameter matrices at the high scale as,
\begin{subequations}\label{eq:GUTbound}
\begin{align}
\bdm^2_{Q,L}&=m^2_{\{Q,L\}0}\dblone+\mathbf{T}_{Q,L}\;, \\
\bdm^2_{U,D,E}&=m^2_{\{U,D,E\}0}[c_{U,D,E}\dblone+R_{U,D,E}\bdf^T_{u,d,e}
\bdf^*_{u,d,e}+S_{U,D,E}(\bdf^T_{u,d,e}\bdf^*_{u,d,e})^2]+\mathbf{T}_{U,D,E}\label{eq:GUTboundsferm}\;,\\
\ba_{u,d,e}&=\bdf_{u,d,e}[A_{\{u,d,e\}0}\dblone+W_{u,d,e}\bdf^\dagger_{u,d,e}\bdf_{u,d,e}+X_{u,d,e}(\bdf^\dagger_{u,d,e}\bdf_{u,d,e})^2]+\mathbf{Z}_{u,d,e}\;,
\label{eq:GUTboundtri}
\end{align}
\end{subequations}
where ${\bf f}_{u,d,e}$ are the superpotential Yukawa coupling matrices {\it in an arbitrary current basis} at the same scale at which the SSB parameters of the model are specified. Here, $c_{U,D,E} =0$ or $1$ is introduced only to allow the facility to ``switch off'' the universal term if desired. We note that \eqref{eq:GUTbound} is a special case of minimal flavor violation \cite{MFV}, when both $\mathbf{T}_{\{Q,L,U,D,E\}}$ and $\mathbf{Z}_{\{U,D,E\}}$ are equal to zero.

As an example of the kinds of effects we observe when introducing one-loop thresholds, consider the well known one-loop RGE invariant relation
\begin{equation}
\frac{\alpha_1}{M_1}=\frac{\alpha_2}{M_2}=\frac{\alpha_3}{M_3}\;,
\end{equation}
 which is valid in models where both the gauge couplings and the gaugino masses unify at the GUT scale.
\begin{figure}\label{gmass}
  \includegraphics[viewport=0 0 665 470,height=.29\textheight,clip]{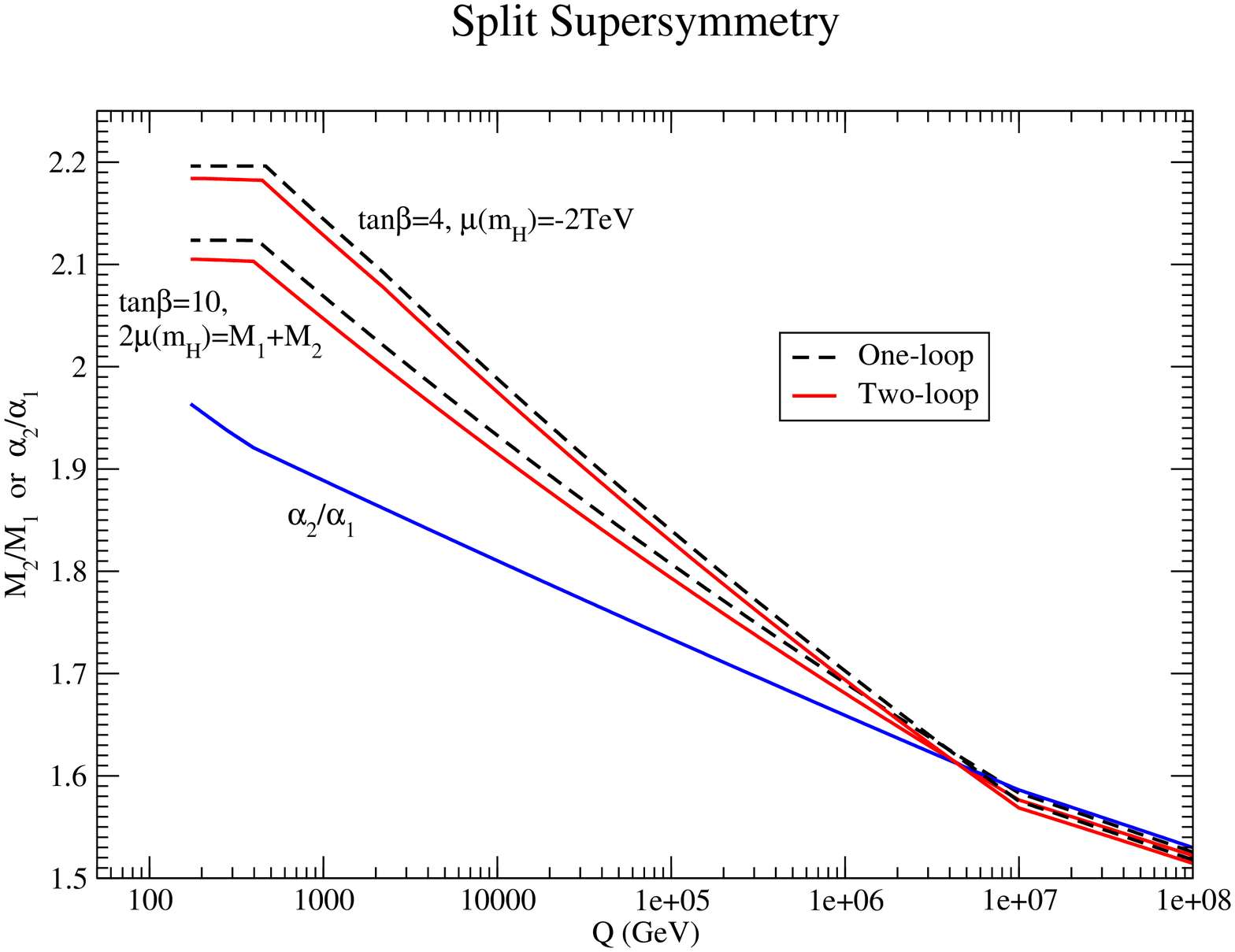}
  \caption{Evolution of the gaugino mass ratio $M_2/M_1$ (solid and dashed lines) along with the two-loop evolution of $\alpha_2/\alpha_1$ (lowest solid line) for a split SUSY model.}
\end{figure}
Figure~\ref{gmass} illustrates the variation of $M_2/M_1$ with renormalization scale, $Q$, in a scenario where the scalars are heavy, at approximately $10^7$~GeV, and the higgsinos and gauginos have masses of the order of the weak scale. We show curves for both the one-loop and two-loop variation of $M_2/M_1$ for two choices of $\tan{\beta}$ and $\mu$, and see that, in the case of low $\tan{\beta}$ and large values of $\mu$, contributions from two-loop terms and threshold corrections to the RGEs cause deviations of the gaugino mass ratio from $\alpha_2/\alpha_1$ by as much as $10\%$.\footnote{This size of this result should only be viewed as a guide, since (in this model) large $\mu$ would be incompatible with relic density measurements and a small value of $\tan{\beta}$ is problematic for EWSB.} We anticipate that increasing the splitting between the scalar and the gaugino/higgsino sector of the theory will drive this figure higher. Lastly, note that the correction introduced by moving to two-loop running is far smaller than that introduced by the thresholds, as a result of the large splitting between SUSY particles.

\section{Flavor Changing top squark decay}

Armed with our method for solving the RGEs to two-loop order, we can revisit the calculation of the two-body flavor changing decay of the lighter stop, $\tilde{t}_1\rightarrow c\tilde{Z}_1$. The rate for this decay was previously studied for light stops by Hikasa and Kobayashi \cite{HK}, who estimated the off-diagonal elements of the up-squark SSB matrix under the approximation that the RGEs could be integrated using a single step.

When we use the full RGE solution we find that the single-step approximation consistently overestimates the width by a factor of between $10-25$. This will clearly have a large effect on the branching ratio in the case that a number of decay modes are of similar order. Indeed, we have shown \cite{RGE2} that in a compressed SUSY scenario \cite{MartinComp} this order of magnitude difference in the rate can significantly change the various branching ratios, and may be even more important in regions of parameter space where the flavor changing decay competes with four-body decays.

Finally, we return to the minimal flavor violation ansatz \cite{MFV} mentioned previously. If we use mSUGRA boundary conditions at the high scale except for
\begin{align}
\label{eq:mqmfv}\mathbf{m}^2_Q=&m^2_{Q0}\left[\dblone+t_u\bdf^*_u\bdf^T_u+t_d\bdf^*_d\bdf^T_d\right]\;,\\
\label{eq:mumfv}\mathbf{m}^2_U=&m^2_{U0}\left[\dblone+R_U\bdf^T_u\bdf^*_u\right]\;,
\end{align}
we can vary $R_U$, $t_u$ and $t_d$ to gauge their influence on flavor violation in the theory.\footnote{Note that our general boundary conditions, \eqref{eq:GUTbound}, are indeed a special case of \eqref{eq:mqmfv} and \eqref{eq:mumfv}, with $t_u=t_d=0$.}

\begin{figure}\label{mfvfig}
  \includegraphics[viewport=0 0 690 515,height=.29\textheight,clip]{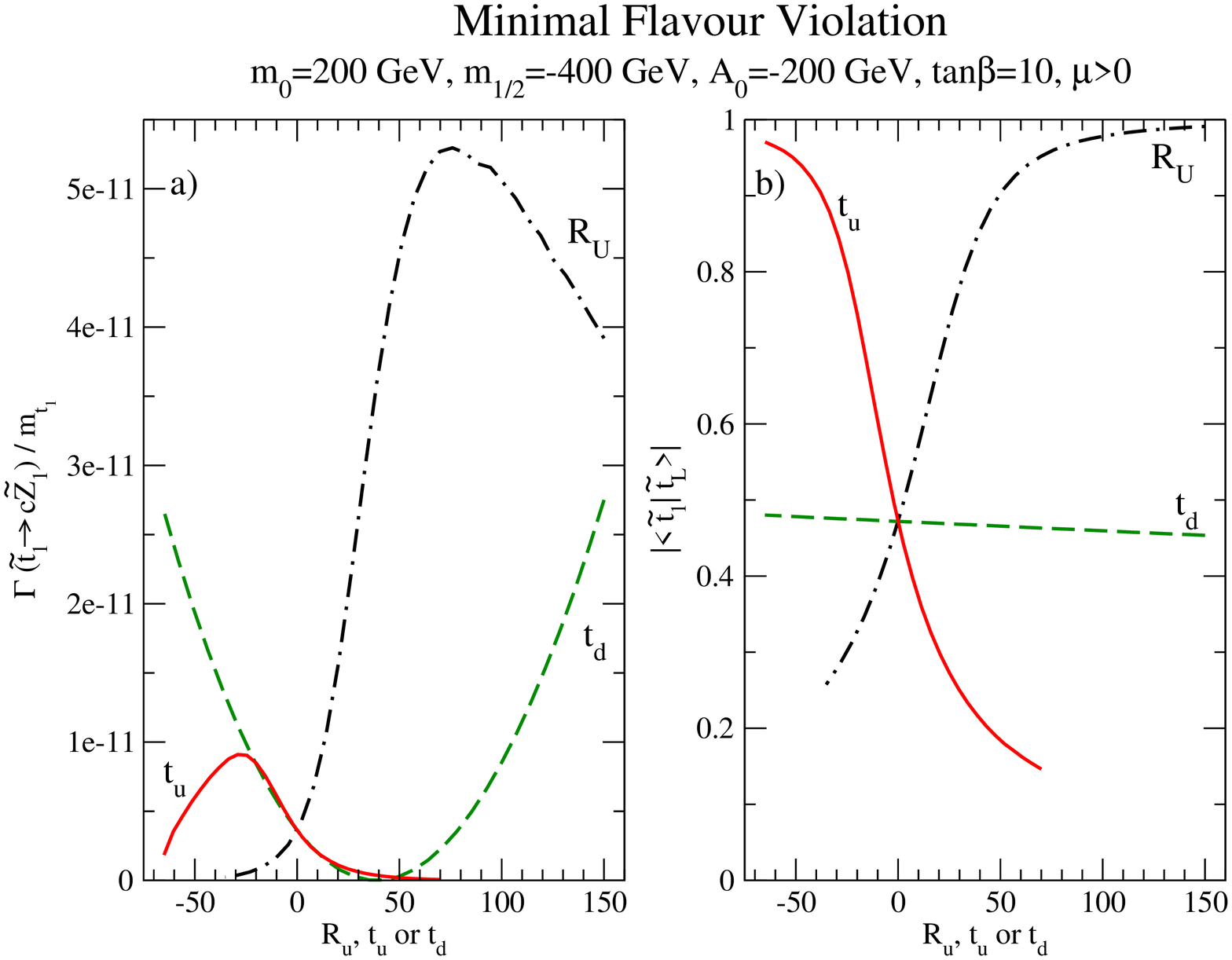}
  \caption{a) Variation of $\Gamma\left(\tilde{t}_1\rightarrow c\tilde{Z}_1\right)/m_{\tilde{t}_1}$ with paramters $R_U$, $t_u$ and $t_d$ as indicated. For each curve, only one MFV parameter labelling that curve has a non-zero value. b) Variation of $|\langle \tilde{t}_1|\tilde{t}_L\rangle|$, the $\tilde{t}_L$ content in the lightest top squark, which would equal $|\cos\theta_t|$ in the absence of any inter-generational mixing.}
\end{figure}

In the left-hand pane of Fig.~\ref{mfvfig}, we plot the variation of the two-body width with the value of either $R_U$, $t_u$ or $t_d$, divided by the stop mass to remove the trivial growth of the width with $m_{\tilde{t}_1}$. Although it would seem that $R_u$ and $t_u$ are introducing new flavor violation, we can see from the right-hand panel of the figure that the widths are roughly tracking the  $\tilde{t}_L-\tilde{t}_R$ ``mixing angle'', which for these parameters is considerably larger than the $\tilde{c}_L-\tilde{t}_R$ mixing term.

On the other hand, the curve for the $t_d\neq0$ case illustrates clear additional flavor violation since the curve in the left-hand pane is not tracking the intra-generation mixing on the right. This is a result of the relatively large contribution to $\mathbf{m}_Q$ from $\bdf_d$, which, in contrast to $\bdf_u$, has large off-diagonal entries.


\begin{theacknowledgments}
This research was carried out in collaboration with Xerxes Tata and was supported in part by a grant from the US Department of Energy.
\end{theacknowledgments}

\end{document}